# Metallic *vs.* Semiconducting Properties of Quasi-One-Dimensional Tantalum Selenide van der Waals Nanoribbons

Fariborz Kargar[1,*], Andrey Krayev[2], Michelle Wurch[1,3], Yassamin Ghafouri[4], Topojit Debnath[5], Darshana Wickramaratne[6], Tina T. Salguero[4], Roger Lake[5], Ludwig Bartels[3], and Alexander A. Balandin[1]

[1] Nano-Device Laboratory (NDL) and Phonon Optimized Engineered Materials (POEM) Center, Department of Electrical and Computer Engineering, University of California, Riverside, California 92521 USA

[2] HORIBA Scientific, Novato, California, 94949 USA

[3] Department of Chemistry and Material Science and Engineering Program, University of California, Riverside, California 92521, United States

[4] Department of Chemistry, University of Georgia, Athens, Georgia 30602 USA

[5]Laboratory for Terahertz and Terascale Electronics, Department of Electrical and Computer Engineering, University of California, Riverside, California 92521, USA

[6]Center for Computational Materials Science, U.S. Naval Research Laboratory, Washington, DC 20375, USA

* Corresponding author: fkargar@ece.ucr.edu ; web-site: https://balandingroup.ucr.edu/

## Abstract

We conducted a tip-enhanced Raman scattering spectroscopy (TERS) and photoluminescence (PL) study of quasi-1D $TaSe_{3-\delta}$ nanoribbons exfoliated onto gold substrates. At a selenium deficiency of $\delta \sim 0.25$ (Se/Ta=2.75,), the nanoribbons exhibit a strong, broad PL peak centered around ~920 nm (1.35 eV), suggesting their semiconducting behavior. Such nanoribbons revealed a strong TERS response under 785-nm laser excitation, allowing for their nanoscale spectroscopic imaging. Nanoribbons with a smaller selenium deficiency (Se/Ta=2.85, $\delta \sim 0.15$) did not show any PL or TERS response. The confocal Raman spectra of these samples agree with the previously-reported spectra of metallic $TaSe_3$. The differences in the optical response of the nanoribbons examined in this study suggest that even small variations in Se content can induce changes in electronic structure, causing samples to exhibit either metallic or semiconducting character. The temperature-dependent electrical measurements of devices fabricated with both types of materials corroborate these observations. The density-functional-theory calculations revealed that incorporation of an oxygen atom in a Se vacancy can result in band gap opening and thus enable the transition from a metal to a semiconductor. However, the predicted bandgap is substantially smaller than that derived from PL data. These results indicate that the properties of van der Waals materials can vary significantly depending on stoichiometry, defect types and concentration, and possibly environmental and substrate effects. In view of this finding, local probing of nanoribbon properties with TERS becomes essential to understanding such low-dimensional systems.

**Introduction**

Recent investigations of two-dimensional (2D) van der Waals materials have revealed new physics and demonstrated potential practical applications.[1–14] Starting with graphene[7–9] and spreading to a wide range of layered van der Waals materials,[10–14]the isolation of individual atomic layers from their respective bulk crystals has led to several breakthrough discoveries. In contrast to graphene or transition metal dichalcogenides (TMDs) that yield quasi-2D samples upon exfoliation, $TiS_3$ and $TaSe_3$ [15–18] yield quasi-one-dimensional (1D) nanostructures. These materials belong to the group of the transition metal trichalcogenides (TMTs) $MX_3$ (where M = various transition metals; X = S, Se, Te). Stoichiometric $TaSe_3$ has a monoclinic structure ($P2_1/m$ space group) with lattice constants $a$=10.402 Å, $b$=3.495 Å, $c$=9.829 Å, and $\beta$=106.26°.[19,20] The atoms form triangular prismatic units with Se atoms at the vertices and Ta in the center, which is repeated along the $b$-axis to form continuous chains (Figure 1 inset). The inter-chain Ta-Ta separation is shorter than intra-chain Ta-Ta separation. The inset in Figure 1 shows the atomic structure of parallel chains revealing a quasi-1D geometry.[19,21,22] The quasi-1D atomic threads are bound weakly in bundles by van der Waals forces and other interchain interactions (Figure S1).

Mechanical exfoliation of the $MX_3$ crystals results not in the 2D layers but rather in the needle-like structures. In some cases, the exfoliated samples are intermediate between the quasi-2D layers and quasi-1D nanowires. In the present study, the exfoliated $TaSe_{(3-\delta)}$ samples had a height in the range from 10 nm to 70 nm, and a width in the range from 100 nm to ~1 μm. For this reason, we refer to them as nanoribbons. In our previous reports, we demonstrated that quasi-1D $TaSe_3$ nanoribbons could sustain a record high current density, $J_B$ exceeding 30 $MAcm^{-2}$, which is an order of magnitude larger than that for the Cu nanowires.[18] The electronic transport characteristics and optical response of such $TaSe_3$ nanoribbons were consistent with their metallic behavior, and in line with early reports of the properties of bulk $TaSe_3$ crystals.[18,23–26] However, some exfoliated quasi-1D nanoribbons of $TaSe_3$ have revealed current-voltage (I-V) characteristics and optical responses more typical for semiconductors. A recent study reported an observation of excitons in exfoliated bundles of $TaSe_3$.[27] It is rather unusual for metals to have excitons due to the high concentration of carriers and corresponding strong screening effects that disfavor exciton formation. These authors argued that dimensional confinement and strong many-body effects in bundles of quasi-1D metallic $TaSe_3$ result in exciton formation.[27]

One should note that even the properties of bulk $TaSe_3$ are not yet fully understood. Because of the low superconducting transition temperature $T_C \sim 2$ K, the bulk electrical characteristics of $TaSe_3$ have not been investigated as thoroughly as of some other TMTs.[19,20,28] Most studies reported that $TaSe_3$ crystals show metallic or semi-metallic behavior down to $T_C$.[20,22,24,29] However, some studies indicated that stress or strain along the long axis can cause an appearance of a semiconducting gap.[30,31] Many reports on $TaSe_3$ crystals do not provide compositional or structural data, *e.g.*, either energy-dispersive spectroscopy (EDX) or X-ray diffraction (XRD). For this reason, it is difficult to assess the composition and quality of the investigated materials. Selenium deficiency, corresponding to $TaSe_{2.8}$, has been observed in prior studies.[32,33] Rather unexpectedly, selenium deficiency in $TaSe_3$ has been reported even in selenium-rich CVT atmospheres.[30,34,35] Prior studies of bulk crystals indicate that the background doping also can adjust the electronic structure of $TaSe_3$.[20,24,28,29,36] For example, as the sulfur content increases in the mixed TMT $Ta(S_xSe_{1-x})_3$, it becomes semiconducting,[37] while $TaSe_3$ with indium impurity exhibits a metal-to-insulator transition.[38,39] It was also reported that copper intercalation into $TaSe_3$ reduces $T_C$ and weakens the charge-density-wave transitions.[40]

Owing to the fast-growing interest in TMTs and other quasi-1D van-der-Waals materials, it is important to understand the fundamental nature of these materials, *i.e.*, metallic vs. semiconducting, and to develop experimental approaches for inspecting the homogeneity or heterogeneity of properties at the nanoscale. In the present study, we show that selenium deficiency, combined with other possible effects, can change the behavior of $TaSe_3$ nanoribbons from metallic to semiconducting. The differences in the optical response and electrical properties of the examined van-der-Waals ribbons suggest that even a small variation in the Se content can induce a change in material's behavior, making it appear more metallic or semiconducting. Our temperature-dependent electrical measurements indicate that the resistance of samples with higher Se deficiency decreases with increasing temperature, revealing a trend characteristic for semiconductors. The resistance of the samples with a composition closer to the stochiometric ideal increases with temperature, in line with a metallic behavior. Our findings may potentially explain some discrepancies in the reported characteristics of exfoliated nanoribbons of $TaSe_3$. Our results attest that a combination of tip-enhanced Raman spectroscopy (TERS), confocal Raman spectroscopy, and photoluminescence (PL) spectroscopy constitutes an effective nanometrology approach for characterization of nanostructures made from van-der-Waals materials. This capability is important for verification of the properties of numerous quasi-1D van der Waals materials predicted by machine learning studies, which are presently being synthesized.[41,42]

## Materials and Methods

The bulk $TaSe_{3-\delta}$ crystals for this study were grown by chemical vapor transport (CVT) or acquired from a commercial vendor (HQ Graphene; also, CVT grown). The details of the CVT synthesis have been reported by some of us elsewhere[18,43,44] and are not reproduced here. The in-house grown and commercial crystals selected for this study have consistent EDS characteristics (see Figure 1). The primary observable difference between these $TaSe_3$ samples is the selenium content. As summarized in Table 1, quantitative EDX characterization provides Se/Ta ratios of 2.85 for CVT-grown crystals and 2.75 for commercial crystals, which correspond to experimental compositions of $TaSe_{2.85}$ and $TaSe_{2.75}$, respectively. Also, the commercial $TaSe_{2.75}$ shows a weak peak which is attributed to oxygen. The crystals were exfoliated to the template-stripped gold substrates, which were detached from the carrier wafer before the $TaSe_{3-\delta}$ exfoliation. We considered two sets of samples – nanoribbons of $TaSe_{2.75}$ and nanoribbons of $TaSe_{2.85}$. The exact dimensions and geometries of the samples were determined by atomic force microscopy (AFM). The AFM, confocal Raman, and TERS characterization were performed using the XploRA-Nano AFM-Raman system (HORIBA Scientific) with the 100× and 0.70 NA side objective inclined at 65º with respect to the normal to the sample's surface. The 785-nm laser excitation with ~400 µW power on the sample surface was used for both the conventional confocal Raman and TERS measurements. The confocal Raman measurements were conducted using DualTwoPass**TM** mode with Access-SNC-Au TERS probes (Applied Nanostructures Inc.). Our TERS measurement procedures have been reported elsewhere.[45,46]

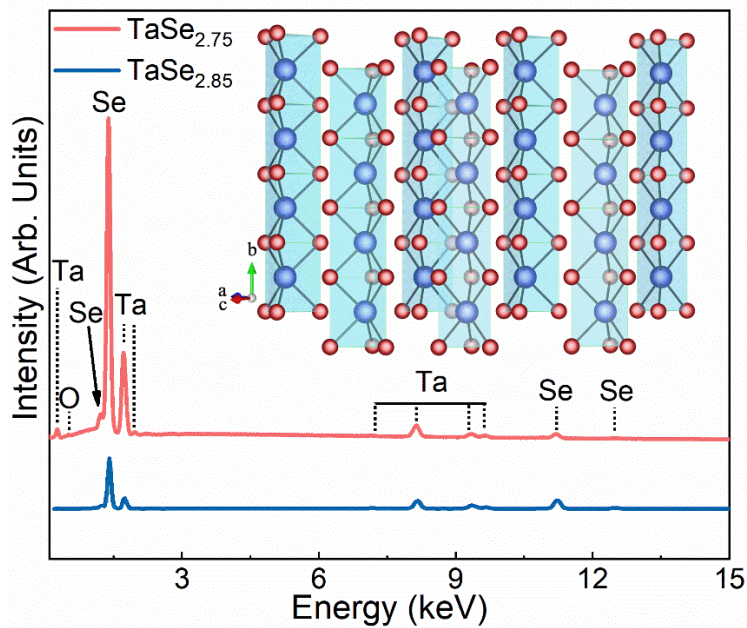

**Figure 1:** Energy dispersive spectra of the synthesized and commercial TaSe$_{3-\delta}$ crystals. The Se/Ta atomic ratio in the commercial sample is 2.75, indicating a larger Se deficiency compared to the CVT-grown crystal with a Se/Ta ratio of 2.85. The inset illustrates the crystal structure of quasi-1D TaSe$_3$, with fused triangular prisms forming parallel chains along the *b*-axis. The red atoms represent Se and the blue ones indicate Ta.

| Table 1: EDX characterization of the TaSe$_{3-\delta}$ crystals used in this study | | | |
|---|---|---|---|
| | Ta (at. %) | Se (at. %) | Ratio Se/Ta |
| Theoretical | 25.00 | 75.00 | 3.00 |
| In-house grown crystal | 26.00 | 74.00 | 2.85 |
| Commercial crystal | 26.69 | 73.31 | 2.75 |

**Results and Discussion**

Figure 2 (a-b) and (c-d), shows the AFM images and scans of two representative exfoliated samples of TaSe$_{3-\delta}$ with $\delta \sim 0.25$ and $\delta \sim 0.15$, respectively. The first set of Raman measurements were conducted using TERS and confocal Raman on the samples shown in Figure 2 (a-d). The spectroscopy results are shown in Figure 2e. In this Figure, the blue and red curves are the confocal Raman and TERS spectra accumulated from TaSe$_{2.75}$. Spectra accumulated by TERS and confocal Raman on samples of TaSe$_{2.85}$ did not show any differences. Therefore, we show only the confocal Raman results of TaSe$_{2.85}$ samples (dark cyan curve, Figure 2e). There are several observations in this Figure that can be listed as follows. Firstly, comparing the spectra of confocal (blue curve) and TERS (red curve) spectra of $TaSe_{2.75}$, one would notice that the intensity of specific peaks, *e.g.*, at 114 cm$^{-1}$ and 263 cm$^{-1}$, in the TERS spectrum is substantially enhanced compared to those observed in the confocal Raman spectrum. The TERS enhancement effect is closely related to the vibrational profile of atoms in those specific phonon modes. The reason for the observed enhancement will be discussed further. On the other hand, none of the exfoliated $TaSe_{2.85}$ samples showed TERS enhancement, *i.e.*, the spectra collected with confocal Raman and TERS system were essentially the same. The second observation is that there are intense Raman peaks at >270 cm$^{-1}$ in TERS spectrum of $TaSe_{2.75}$ (red curve) which are absent or very weak in the confocal Raman spectrum of the same crystal. These peaks are attributed to the second- or multi-order Raman scattering processes. These high wavenumber peaks are absent in the data accumulated for $TaSe_{2.85}$ (cyan curve). A third observation is that the Raman spectra of $TaSe_{2.75}$ samples are accompanied by an intense photoluminescence background at higher energy, which is not seen in the spectrum collected for $TaSe_{2.85}$. This observation suggests that samples with larger Se deficiency possess semiconductor characteristics whereas samples with close to stoichiometric composition are metallic. We will address the photoluminescence in detail. Fourth, although the Raman spectra of the samples look superficially different in the range between 100 cm$^{-1}$ to 300 cm$^{-1}$, in fact both agree well with the calculations, with only the intensities of the peaks differing. $TaSe_3$ has many Raman signatures, with closely-packed Raman peaks that will be discussed in more details below. We focus here on the TERS data of $TaSe_{2.75}$ given its more intense and well-defined spectral features.

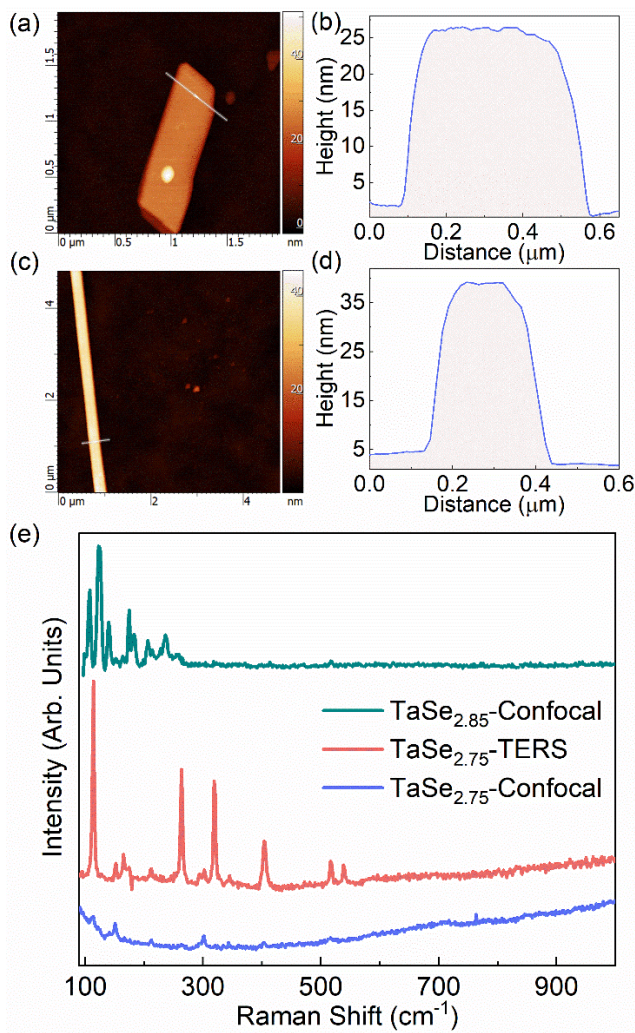

**Figure 2:** AFM topography analysis of (a-b) TaSe$_{2.75}$ and (c-d) TaSe$_{2.85}$ nanoribbons. (e) TERS and confocal Raman spectra accumulated for the samples shown in panels (a, c). Exfoliated samples with higher Se deficiency exhibit TERS enhancement and second-order Raman modes with strong photoluminescence background (red and blue curves) whereas TaSe$_{2.85}$ samples exhibit neither of these characteristics.

Figure 3a shows the AFM thickness analyses of the sample used for TERS experiments. The AFM image is presented in the Supplementary Figure S2. Figure 3b shows the intensity distribution of the most dominant Raman peaks at 265 cm$^{-1}$ (red color), 320 cm$^{-1}$ (blue color) and 405 cm$^{-1}$ (green color), respectively. The data were accumulated under 785-nm laser excitation at 400 μW power on the sample surface, and 1 second per pixel integration time. The map repeats the topography image of the exfoliated TaSe$_3$ sample shown in Supplementary Figure S2 with high fidelity. There is a sharp drop in the signal intensity at the edges of the crystal, which indicates the lack of significant far-field Raman contribution to the collected spectra. Interestingly, TERS spectra averaged over the darker stripes in the TERS map showed not only an overall decrease in the signal intensity of the intense bands at 265 cm⁻¹, 320 cm⁻¹ and 405 cm⁻¹ but also a disproportionate decrease in the intensity and a slight red shift of the 166 and 540 cm⁻¹ bands. Such spectral deviations suggest the presence of defects in the sample that do not show up in topography. Figure 3c shows the TERS of the same sample in the spectral range of 100–600 cm⁻¹. The data represents the averaged intensity of the peaks over the bright (blue curve) and dim (red curve) spots. The green curve is the cumulative fitting over the experimental data points using individual Lorentzian functions. In this spectrum, 14 Raman peaks are represented: at 127, 142, 152, 166, 174, 213, 264, 294, 302, 320, 349, 405, 517, and 539 cm⁻¹. A low-intensity peak at 187 cm⁻¹ could not be fit using the Lorentzian or other functions due to its low intensity. Furthermore, a series of confocal Raman measurements were conducted using special notch filters with the cut-off at ~60 cm⁻¹. In this way, two peaks at ~75 cm⁻¹ and 82 cm⁻¹ were identified (Supplementary Figure S3).

As illustrated in Figure 4a and reported by some of us previously,[47] the maximum energy of the optical phonons along all the high-symmetry directions fall below ~270 cm⁻¹ [see Ref [47] and Figure 4a] and those Raman peaks observed at higher frequencies than 270 cm⁻¹ belong to the multi-order Raman scattering processes. These high-order Raman peaks are a measure of the phonon density of states (PDoS).[48,49] Momentum conservation in two-phonon scattering is satisfied when $q_1 \pm q_2 \sim 0$ in which $q_i, i = 1,2$ is the wavevector of the specific phonon mode contributing to the scattering and plus and minus signs represent combination and difference modes. Therefore, the momentum conservation restrictions of single-phonon scattering do not apply in multi-scattering processes and basically, phonons satisfying the above condition may appear in the Raman spectrum. The details surrounding the selection rules for higher-order Raman scattering processes in TaSe₃₋δ is beyond the scope of this investigation. However, it is well-established that higher-order Raman scattering processes can be activated by the inherent local defects, edges, or impurities in crystalline materials.[48–50] Examples of such multi-phonon processes have been reported in a variety of material systems, including conventional materials like doped and nanocrystalline bulk silicon as well as low-dimensional, novel materials like graphene and black phosphorus. For instance, in graphene, a phonon with D and D' Raman bands is turned on as a consequence of phonon-electron interaction activated by local impurities or edges.[50] In fact, appearance of higher-order Raman bands in TaSe$_{2.75}$ with larger Se deficiency and their absence in TaSe$_{2.85}$ indicates the role of defects and a possible change in the electronic band structure and optical properties. As seen in Figure 3c, at higher wavenumbers >600 cm$^{-1}$, the Raman spectra are accompanied with a strong background (the colored area) attributed to the PL tail. In all experiments, the presence of the PL tail was accompanied consistently by the appearance of higher-order Raman modes, further indicating the role of Se deficiency or other defects in changing this material's electronic properties from metallic to semiconductor.

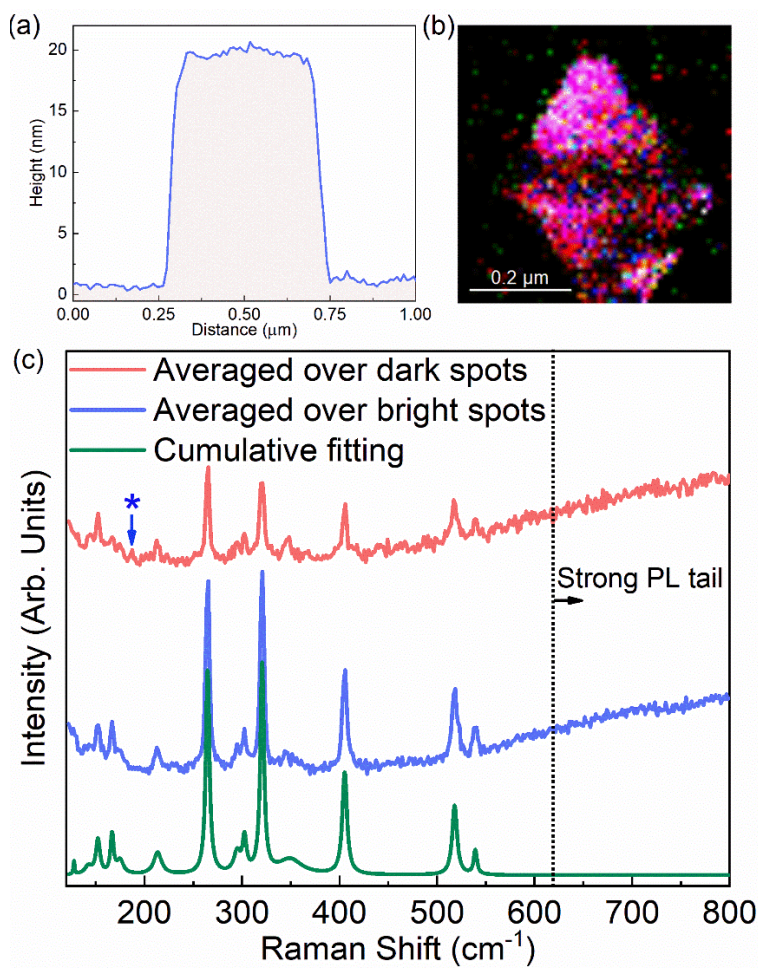

**Figure 3:** (a) AFM topography analysis of the TaSe$_{2.75}$ sample used in TERS measurement. (b) TERS contour map presenting the intensity distribution of Raman dominant peaks at 265 cm$^{-1}$, 320 cm$^{-1}$ and 405 cm$^{-1}$, respectively. (c) TERS spectra averaged over the bright (blue spectrum) and dim (red spectrum) areas in the TERS map presented in panel (b). The green curve shows the cumulative fitting over the experimental data using individual Lorentzian functions. The peak labeled with "*" belong to $A_g$ vibrational symmetry and could not be fitted due to its low intensity.

To rationalize our experimental Raman data, we calculated the phonon dispersion and PDOS of stoichiometric $TaSe_3$ using the density functional theory (DFT). The results are presented in Figure 4a. The red symbols represent our experimental Raman peaks for $TaSe_{2.75}$. As one can see, the experimental data points agree well with the theoretical calculations and the previously reported values.[47,51] Table S1 in the Supplementary Information lists the spectral positions of the Raman peaks, theoretical calculations for the Brillouin zone center ($\Gamma$), and the previously reported data in the literature. The unit cell of $TaSe_3$ has 16 atoms (see Figure S1 (a)) with 45 optical phonon branches. This large number results in rich Raman response, illustrated in Figure 3c. The vibrational symmetry of the atomic chain and the crystal are represented by: $\Gamma_{TaSe_3} = 8A_u + 8B_g + 16B_u + 16A_g$, where $A_g$ and $B_g$ modes are Raman active.[51] The vibrational symmetry of Raman active modes is also included in Table S1. We observed experimentally TERS enhancement of the Raman peaks at 114 cm$^{-1}$ and 263 cm$^{-1}$ (see Figure 2e). These two peaks have $A_g$ symmetry with strong displacement of atoms in the *a-c* plane (see Figure 4b). In TERS experiments, a strong electric field of the localized surface plasmons between the apex of the metallic gold tip and sample's surface intensifies Raman peaks owing to 4$^{th}$ order dependence of the Raman intensity, $I$, on the magnitude of the local electric field ($E$), *i.e.,* $I \sim E^4$. The direction of the induced localized electric field lies along the normal to the sample's surface. Consequently, the Raman-active modes with the large atomic vibrational profiles and polarizability along the confined electric field are enhanced preferentially.

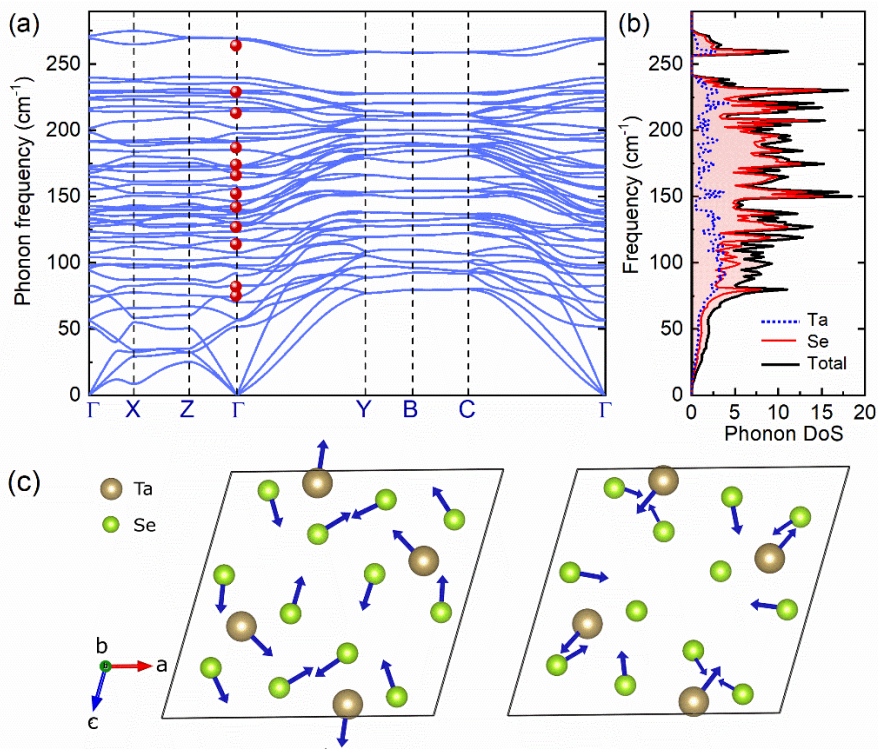

**Figure 4:** (a) Phonon band structure, and (b) phonon density of states of stoichiometric TaSe$_3$. The red symbols are the experimental Raman peaks accumulated form TaSe$_{2.75}$. (c) vibrational profiles and symmetry of the Raman bands at 114 cm$^{-1}$ (left) and 263 cm$^{-1}$ (right) with strong displacement profile of atoms in the *a-c* plane. The intensity of these modes is preferentially enhanced in TERS measurements.

The increasing background in TERS spectra collected over the 20 nm thick nanoribbon suggest the presence of a PL peak tail (see spectra in Figure 2e for >600 cm$^{-1}$). To determine the exact position of this PL peak, and to confirm that other exfoliated TaSe$_{2.75}$ samples also show TERS response, we performed TERS and tip enhanced PL (TEPL) imaging of several nanoribbons and small samples of different geometries. We used the lower density grating in the Raman instrument to expand the covered spectral range in a single measurement. A broad PL peak with a maximum at approximately ~922 nm (1.34 eV) was present in both the TERS and conventional confocal Raman spectra collected over the same samples. To verify this observation, we conducted PL mapping on a 70 nm thick sample with larger lateral dimensions. The AFM image and thickness analysis of the sample are presented in the Supplementary Figure S4 and the inset of Figure 5, respectively. The results of PL measurements are shown in Figure 5. The PL peak at 922 nm is consistently present for the samples with larger Se deficiency ($\delta$~0.25). The strong PL response of this sample allowed us to use a short integration time of 20 ms per pixel and to collect a high pixel density PL map, which followed the topography image with high fidelity (Figure S5). In samples with lower Se deficiency ($\delta \sim 0.15$), the PL peak was not observed. This observation suggests the material's electronic characteristic can undergo substantial changes with varying Ta/Se proportions.

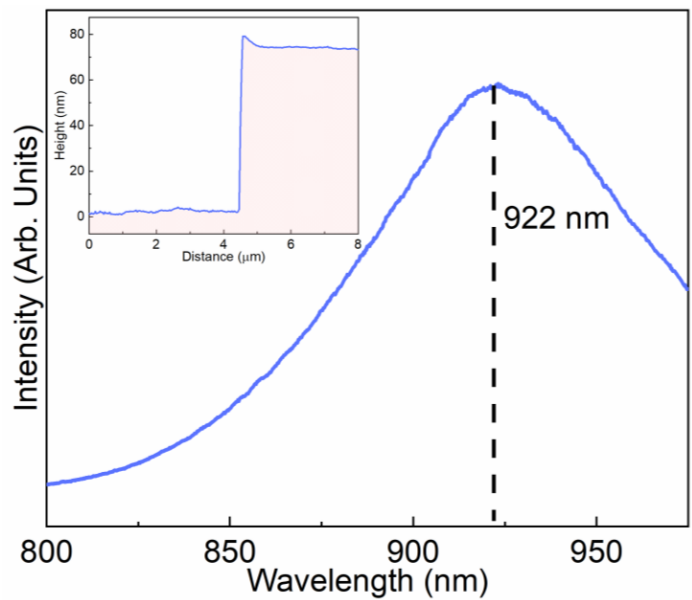

**Figure 5:** PL spectrum averaged over the sample surface indicating the PL maximum at 922 nm (1.34 eV). The inset presents the AFM scan of the sample showing its thickness.

These results establish that TERS enhancement, the appearance of the high-order Raman modes, and the PL tail and peak can be used as tools to distinguish semiconducting vs. metallic samples of exfoliated $TaSe_{3-\delta}$. As discussed above, most of the reported studies indicate metallic behavior by $TaSe_3$ in both transport and optical measurements.[23–25] The examined CVT-grown $TaSe_{2.85}$ nanoribbons did not show any PL or TERS enhancement response whereas $TaSe_{2.75}$ exhibited both. For this reason, we hypothesized the differences in optical spectra can be related to the degree of selenium deficiency. The samples with the composition closer to the stoichiometric $TaSe_3$ are more likely to be metallic. In order to verify this hypothesis, we fabricated devices with $TaSe_{3-\delta}$ channels and measured their I-V characteristics as a function of temperature. The results of such measurements for $TaSe_{2.75}$ and $TaSe_{2.85}$ are presented in Figure 6 (a-b). The insets in the figure show the SEM images of the tested devices. The resistance of the devices fabricated with TaSe$_{2.75}$ increases with temperature, whereas it decreases for TaSe$_{2.85}$ devices as the temperature increases. These trends confirm that the samples with larger Se deficiency reveal semiconducting behavior.

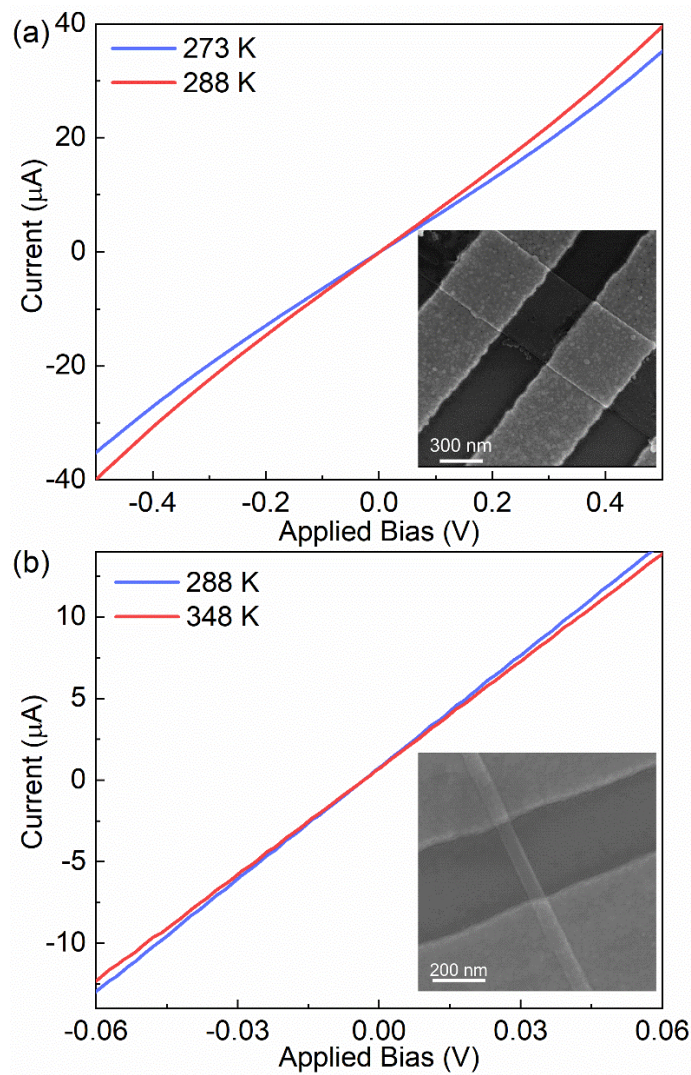

**Figure 6:** I-V characteristics of devices with (a) TaSe$_{2.75}$, and (b) TaSe$_{2.85}$ channels shown at two different temperatures. The electrical resistance of the devices with the TaSe$_{2.75}$ and TaSe$_{2.85}$ channels decreases and increases with temperature rise, respectively. This indicates a transition from the semiconducting to metallic conduction depending on the degree of Se deficiency.

To explore the possibility of band-gap opening due to composition and defects, we theoretically investigated the effect of Se vacancies and Se/O substitution. We calculated the density of states (DOS) of the stoichiometric bulk $TaSe_3$ crystal, the Se deficient $TaSe_{2.75}$ containing Se vacancies, and an O-substituted structure $TaSe_{2.75}O_{0.25}$ using Density Functional Theory (DFT).[52,53] The stoichiometric $TaSe_3$ unit cell has 4 Ta atoms and 12 Se atoms. To create the Se deficient structure of $TaSe_{2.75}$, we removed one Se atom from the stoichiometric unit cell. Depending on the position of removed Se atom, two different configurations were considered, interchain and intrachain vacancies, as shown in Figure 7. In the interchain configuration, a Se atom that connects two Ta atoms from adjacent chains is removed. In the intrachain configuration, a Se atom, connecting two Ta atoms of same chain, is removed. Due to the well-known oxidation of metal chalcogenides, we considered two more interchain and intrachain configurations, in which the Se vacancy is filled by oxygen.

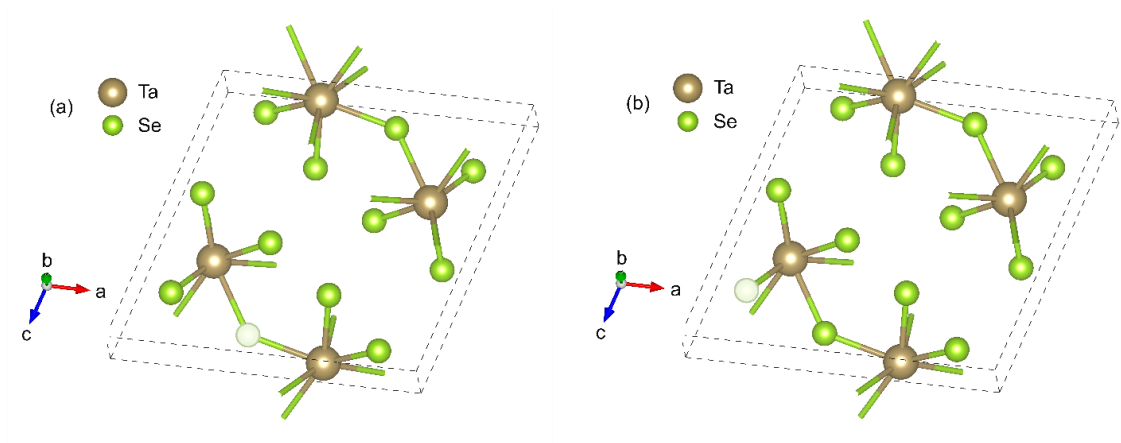

**Figure 7:** (a) Interchain and (b) Intrachain vacancy configuration of $TaSe_{2.75}$. The twelve atoms within the unit cell are shown. The transparent spheres represent Se vacancies or O-substituted Se atoms.

Structure relaxation of all configurations were performed using the Vienna *ab initio* Simulation Package (VASP)[54] and the Python based atomic simulation environment (ASE).[55] We used the Perdew-Burke-Ernzenhof (PBE) exchange correlation (xc) functional[56] and a plane wave basis set with an energy cutoff at 500 eV to calculate the ground state energy of these materials. The DFT-D2 method proposed by Grimme[57] is included in these calculations to account for the van der Waals interaction. The structures were relaxed until the maximum force on all individual atoms was less than 0.1 meV/A. The total energies of the four structures listed in the Table 2, show that the intrachain vacancy or O-substitution is more stable than its corresponding interchain one.

| Table 2: Total energy (eV) of four atomic configurations | | | | |
|---|---|---|---|---|
| Structure | Se - Vacancy | | Oxygen substituted | |
| | Interchain | Intrachain | Interchain | Intrachain |
| Total Energy (eV) | -125.489 | -127.334 | -138.474 | -138.512 |

To investigate the effect of vacancies and O-substitution on the electronic band structure, we used the Heyd-Scuseria-Ernzerhof (HSE06) hybrid functional[55] with and without spin orbit coupling (SOC), since the SOC of both Ta and Se is large. The Brillouin zone was sampled by a 6x14x6 Monkhorst-Pack k-point grid. Figure 8 shows the HSE06 and HSE06-SOC DOS plots resulting from interchain and intrachain configurations for both the Se deficient and oxygen substituted structures. The left column shows the DOS with the Se vacancy, and the right column shows the DOS with O-substitution. The top row is without SOC and the bottom row is with SOC. The dashed black curves show the DOS of the stoichiometric structure. Comparing the DOS of the stoichiometric structure calculated with and without SOC, we see that the effect of SOC is to reduce the DOS of the stoichiometric structure near the Fermi level. In the absence of SOC, there are several bands that cross near the Fermi level;[47] taking SOC into account, these band crossings become anti-crossings, which reduces the spectral weight near the Fermi level. Intrachain Se vacancies increases the DOS near the Fermi level with or without SOC. The effect of interchain vacancies depends on the presence or absence of SOC. Without SOC, interchain vacancies reduce the DOS, whereas with SOC, they cause a slight increase in the DOS at the Fermi level.

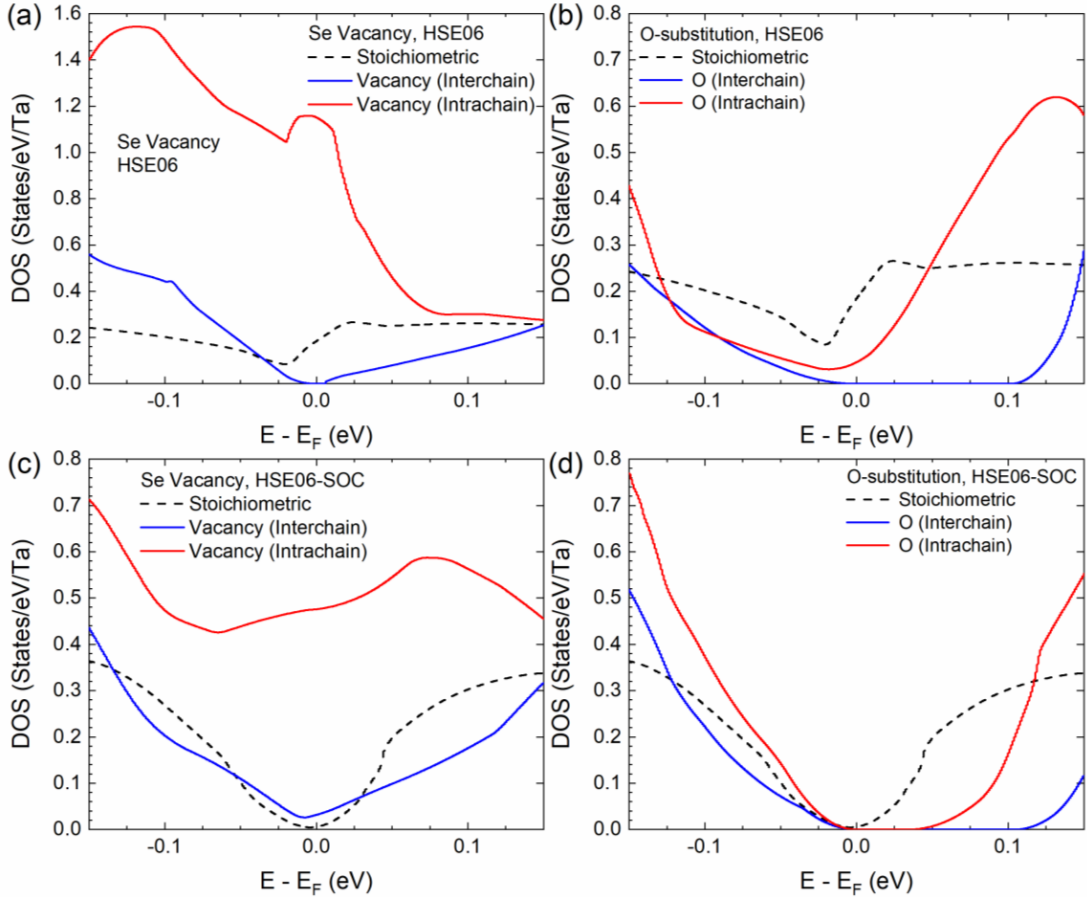

**Figure 8:** Electronic density of states for (a) Se deficient structure TaSe$_{2.75}$, (b) oxygen substituted structure TaSe$_{2.75}$O with HSE only. Figure (c) and (d) shows the dos when both HSE and SOC are taken into consideration.

As one can see from the simulations results, we do not find a compelling case for Se vacancies leading to the energy bandgap extracted from the PL data. The effect of Se/O-substitution gives a more consistent picture independent of the presence or absence of SOC. In all cases, O substitution results in a reduction of DOS near the Fermi level. In the presence of SOC, O substitution results in a small gap around the Fermi level with a larger gap for interchain substitution. The size of the gap, however, is relatively small, between 0.05 and 0.1 eV, which is approximately an order of magnitude less than the experimentally observed PL peak. The simulations show that the vacancies alone, with the concentrations in the considered range, cannot explain the observed optical response. Oxygen substitution can create an energy band gap but its magnitude remains relatively small. Interestingly, oxygen substitution for S atoms was reported to open up a bandgap in another material such as $TiS_{2-x}O_x$. It was found that the band gap monotonically increases with oxygen concentration.[58] In addition to the deviation from the stoichiometry and oxygenation, one should assume possible influence of other defects, environmental and substrate effects, *e.g.* strain, to explain the experimentally observed differences in properties.

## Conclusions

We conducted a tip-enhanced Raman scattering spectroscopy and photoluminescence study of nanoribbons exfoliated from two quasi-1D $TaSe_{3-\delta}$ samples, with Se deficiency of $\delta \sim 0.15$ and $\delta \sim 0.25$. The $TaSe_{2.75}$ nanoribbons exhibited strong, broad PL peak centered around 920 nm (1.35 eV) and TERS response under 785-nm laser excitation, consistent with semiconducting behavior. The $TaSe_{2.85}$ nanoribbons showed neither PL nor TERS response, in agreement with the metallic characteristic of stochiometric $TaSe_3$. The confocal Raman spectra of ribbons of both types agreed well with the previously reported spectra of metallic $TaSe_3$. The temperature-dependent electrical measurements on two-terminal devices fabricated with nanoribbons of $TaSe_{2.85}$ and $TaSe_{2.75}$ indicated semiconducting and metallic trends, in line with the optical studies. The density-functional theory calculations of the electronic band structure suggested that oxygen substitution rather than Se vacancies can produce band gap opening in this system. Oxygen substitution results in a relatively small bandgap opening in the range of 0.05 to 0.1 eV. The calculated bandgap is an order of magnitude smaller than that experimentally observed in PL measurements. The differences in the optical response and electrical transport of the examined nanoribbons suggest that even small variations in Se content can induce changes in this material's behavior, making it appear metallic or semiconducting. This observation may explain some discrepancies in the earlier reported transport characteristics of exfoliated nanoribbons of $TaSe_3$. Our results also attest that a combination of TERS and PL spectroscopy constitutes a powerful nanometrology tool for the characterization of nanostructures made from van der Waals materials.

## Acknowledgements

The work at UCR was supported, in part, by the National Science Foundation (NSF) program Designing Materials to Revolutionize and Engineer our Future (DMREF) *via* a project DMR-1921958 entitled Collaborative Research: Data Driven Discovery of Synthesis Pathways and Distinguishing Electronic Phenomena of 1D van der Waals Bonded Solids. A.A.B., L.B. and F.K. acknowledge useful discussions with Evan Reed (Stanford University).

**Contributions**

A.A.B. coordinated the project and contributed to the data analysis. F.K. and A.K. conceived the idea of the study. Y.G. synthesized bulk crystals and conducted material characterization. T.T.S. supervised material synthesis and contributed to data analysis. A.K. performed TERS measurements and contributed to data analysis. M.W. fabricated the test structures and conducted electrical measurements. L.B. supervised nanofabrication and contributed to data analysis. T.D. and D.W. performed *ab initio* computational studies. R.L. conducted theoretical analysis. All authors contributed to writing and editing of the manuscript.

# Supplementary Information

# Metallic *vs.* Semiconducting Properties of Quasi-One-Dimensional Tantalum Selenide van der Waals Nanoribbons

Fariborz Kargar[1,†], Andrey Krayev[2], Michelle Wurch[1,3], Yassamin Ghafouri[4], Topojit Debnath[5], Darshana Wickramaratne[6], Tina T. Salguero[4], Roger Lake[5], Ludwig Bartels[3], and Alexander A. Balandin[1]

[1] Nano-Device Laboratory (NDL) and Phonon Optimized Engineered Materials (POEM) Center, Department of Electrical and Computer Engineering, University of California, Riverside, California 92521 USA

[2] HORIBA Scientific, Novato, California, 94949 USA

[3] Department of Chemistry and Material Science and Engineering Program, University of California, Riverside, California 92521, United States

[4] Department of Chemistry, University of Georgia, Athens, Georgia 30602 USA

[5] Laboratory for Terahertz and Terascale Electronics, Department of Electrical and Computer Engineering, University of California, Riverside, California 92521, USA

[6] Center for Computational Materials Science, U.S. Naval Research Laboratory, Washington, DC 20375, USA

---

[†] Corresponding author: fkargar@ece.ucr.edu ; web-site: https://balandingroup.ucr.edu/

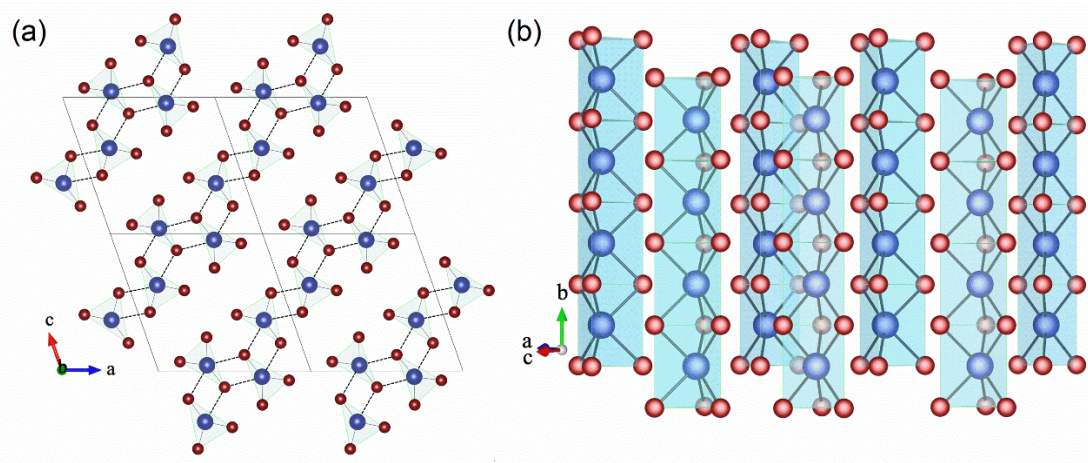

**Figure S1:** (a-b) Schematic of the crystal structure of quasi-1D TaSe$_3$ and the triangular prisms forming parallel van der Waals chains from different views. The red and blue circles represent Ta and Se atoms.]

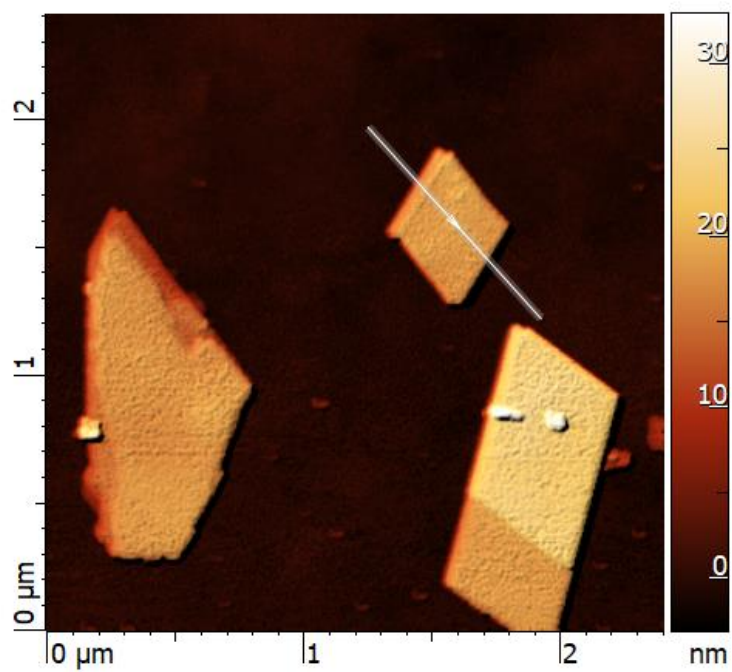

**Figure S2:** AFM image of the sample used in Figure 3a.

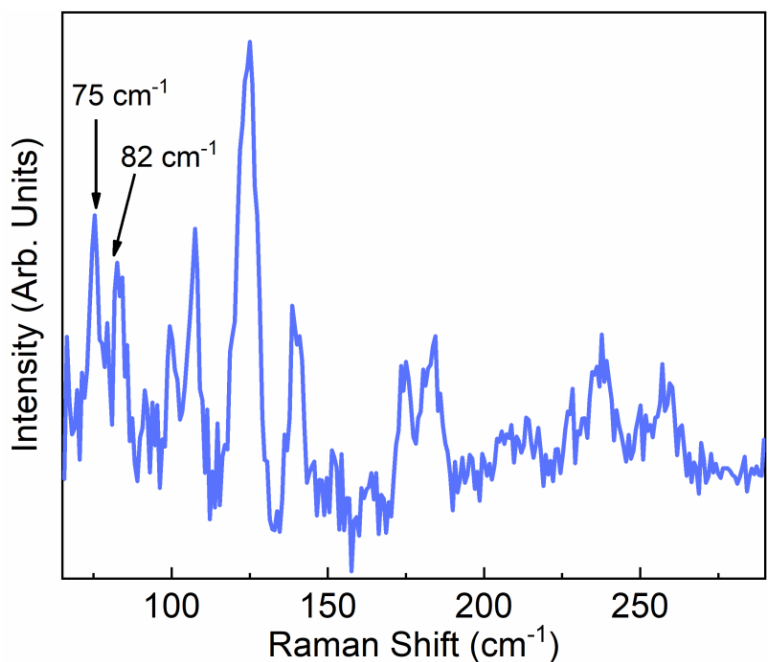

**Figure S3: (a-b)**: Low-wave-number Raman experiments showing the low-frequency Raman modes at 75 cm⁻¹ and 82 cm⁻¹.

| Table S1: Raman active modes of TaSe3 in the range of 65 cm⁻¹ to 270 cm⁻¹ | | |
|:---:|:---:|:---:|
| Experiment | Theory | Symmetry |
| 75 | 74.5 | Bg |
| 82 | 82.7 | Ag |
| - | 107.5 | Ag |
| 114 | 116.7 | Ag |
| - | 123.4 | Bg |
| 127 | 127.6 | Bg |
| - | 133.4 | Bg |
| - | 138.1 | Bg |
| 142 | 140.1 | Bg |
| - | 146.4 | Bg |
| 152 | 155.4 | Ag |
| 166 | 168.0 | Ag |
| 174 | 173.2 | Bg |
| 187 | 185.3 | Ag |
| - | 191.7 | Bg |
| | 201.8 | Ag |
| 213 | 216.8 | Ag |
| - | 224.9 | Ag |
| 229 (very weak) | 228.3 | Ag |
| - | 239.9 | Ag |
| 264 | 269.4 | Ag |

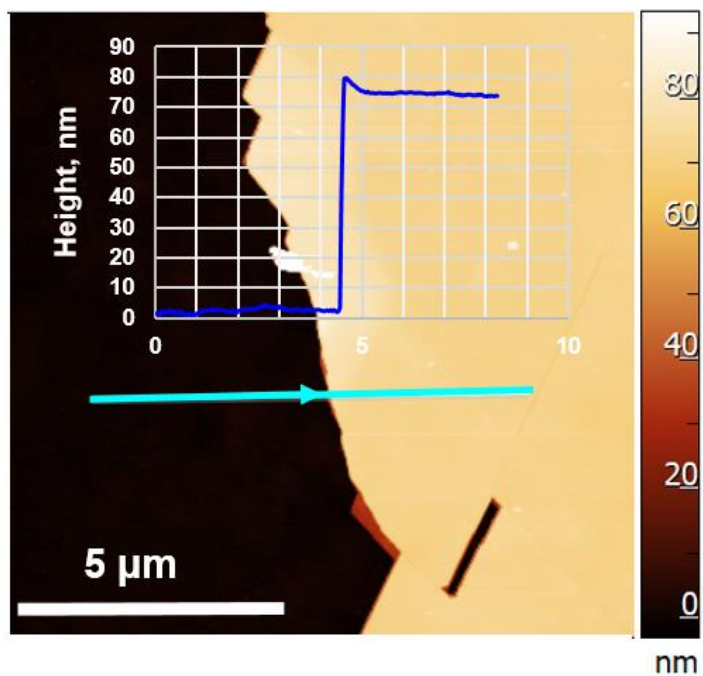

**Figure S4:** (a) AFM topography image (12 µm × 12 µm) of a large TaSe$_{2.75}$ crystal on a template stripped gold substrate. In the insert-section graph along the cyan line showing that the height of the crystal was about 70 nm.

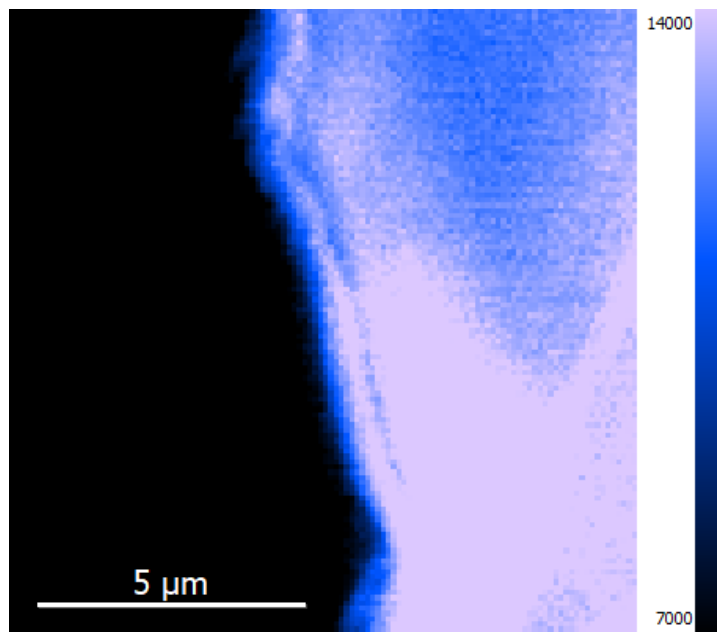

Figure S5: Confocal PL map of the crystal used in PL measurement.

## TERS measurement procedure in DualTwoPass$^{TM}$ mode

As the TERS signal gets strongly enhanced when the TERS probe is in direct contact with the sample, TERS measurements presented in current study were performed using proprietary DualTwoPass imaging mode. The measurement procedure was the following: in the beginning of each line of the TERS map the AFM was operating in AC (tapping) mode. Then the feedback was switched to the contact mode with normal force exerted on the samples of about 100 nN, and this contact mode feedback was maintained for the duration of the line. A full Raman spectrum was collected in each pixel along the line with integration time ranging from a few hundred milliseconds to a few seconds per pixel. At the end of each line the feedback was switched to AC (tapping) operation, the sample was moved to the beginning of the next line, switched to contact mode and so forth. Since in every pixel of the map a full spectrum was collected, TERS maps could be rendered as the intensity or the peak position maps of up to three different peaks.

When regular confocal Raman signal was to be collected along with the TERS signal, after the contact mode pass, the same line was scanned again in AC (tapping) mode when the Raman signal was of purely far field nature. Thus, we collected simultaneously two maps- a far field map and the map containing both the TERS contribution and (usually much weaker) the far field signal. Subtracting the far field map from the combined map we could obtain the pure TERS (near field) map.